\documentclass[10pt,a4paper]{article}
 \usepackage{graphicx} % Include figure files

\begin{document}

\begin{center}

{\bf STRANGE MULTIBARYON STATES WITH $\Lambda$ HYPERON SYSTEMS IN pA
COLLISION AT 10 GeV/c }

\vskip 5mm P.Zh.Aslanyan$^{\dag,\ddag}$

\vskip 5mm

{\small {\it $^\dag$ Joint Institute for
Nuclear Research, 141980 Dubna, Russia}} \\
{\small {\it $^\ddag$ Yerevan State of University}}
\\

PACS{\small {\it:14.20.Jn, 14.40.Aq, 25.80.Nv, 25.80.Pw, 14.20.Gk,
14.40.Ev, 14.20.Pt}}

\end{center}

\vskip 5mm

\centerline{\bf Abstract} Experimental data from the 2m propane
bubble chamber have been analyzed for exotic baryon states search. A
number of peculiarities were found in the effective mass spectra of:
$\Lambda \pi^+$($\Sigma^{*+}$(1382),PDG), $\Lambda p$ and $\Lambda p
p$ subsystems. A few events detected on the photographs of the
propane bubble chamber exposed to a 10 GeV/c proton beam, were
interpreted as S=-2 $H^0$ light($<M_{(\Lambda \Lambda)}$) and heavy
$H^{0,+}$. New event, detected on the photographs of the propane
bubble chamber was interpreted as  heavy $H^+(2488)$ dibaryon by two
weak decay channels of $K^-pp$ or $ \Sigma^+ p\pi^-$.

\vskip 10mm

\section{\label{sec:intro}Introduction}

Already back in 1977 Jaffe\cite{Jaffe}, using the bag model in which
confined colored quarks and gluons interact as in perturbative QCD,
suggested the existence of multi-quark states, glueballs and
hybrids, but until now none is established. Very recently, the
existence of discrete nuclear bound states of $\overline{K}^0$p has
been predicted with phenomenological Kaonic Nuclear Cluster (KNC)
model which is based on the experimental information on the
$\overline{K}^0$N scattering lengths, kaonic hydrogen atom and the
$\Lambda^*(1405)$ resonance\cite{knc}.

 A number of peculiarities were found in the effective mass spectrum
 of $\Lambda \pi^+$(from PDG),$\Lambda p$ and $\Lambda p p$ systems and
some above peaks were conformed with results from FOPI,GSI and E471,KEK.\\
 A few events detected on the photographs of the propane bubble chamber
exposed to a 10 GeV/c proton beam, were interpreted as S=-2 $H^0$
light($<M_{(\Lambda \Lambda)}$) and heavy $H^{0,+}$
dibaryons\cite{H}and were identified by hypothesis in weak decay
channels of $\Sigma^- p$,$\Lambda p \pi^0$,$\Lambda p
\pi^-$,$\Sigma^+ p \pi^-$ and $K^- pp$ .
%% To insert figure (with the help of epsf.sty)

\section{Method}

The full experimental information of more than 700000 stereo
photographs or $10^6$ p+propane inelastic interactions are used to
select the events with $V^0$ strange particles. The masses of the
observed 8657-events with $\Lambda$ hyperon  4122-events with
$K_s^0$ meson  are consistent with their PDG values\cite{v0}. The
experimental total cross sections are equal to 13.3 and 4.6 mb for
$\Lambda$ and $K_s^0$ production in the p+C collisions at 10 GeV/c.
From published article  one can see that the experiment is
satisfactorily described by the FRITIOF model.The experimental
$\Lambda /\pi^+$ ratio in the pC reaction is approximately two times
larger than this ratio  from  pp reactions or from simulated pC
reactions by FRITIOF model  at the same energy \cite{v0}. The
resonance with similar decay properties $\Sigma^{*+}(1382)\to\Lambda
\pi^+$ registered as test for this method (Fig. \ref{lpip}a). The
decay width is equal to  $\Gamma$= 40 MeV/$c^2$. $\Delta M/M =0.7$
in range  of $\Sigma^*(1382)$ invariant mass.  Just the cross
section of  $\Sigma^*(1382)$ production (540 simulated events) can
estimated by FRITIOF model which is approximately equal to 1 mb for
p+C interaction.

 \section{($\Lambda, p$)  spectra}

\begin{figure}[ht]
\includegraphics[width=45mm,height=70mm]{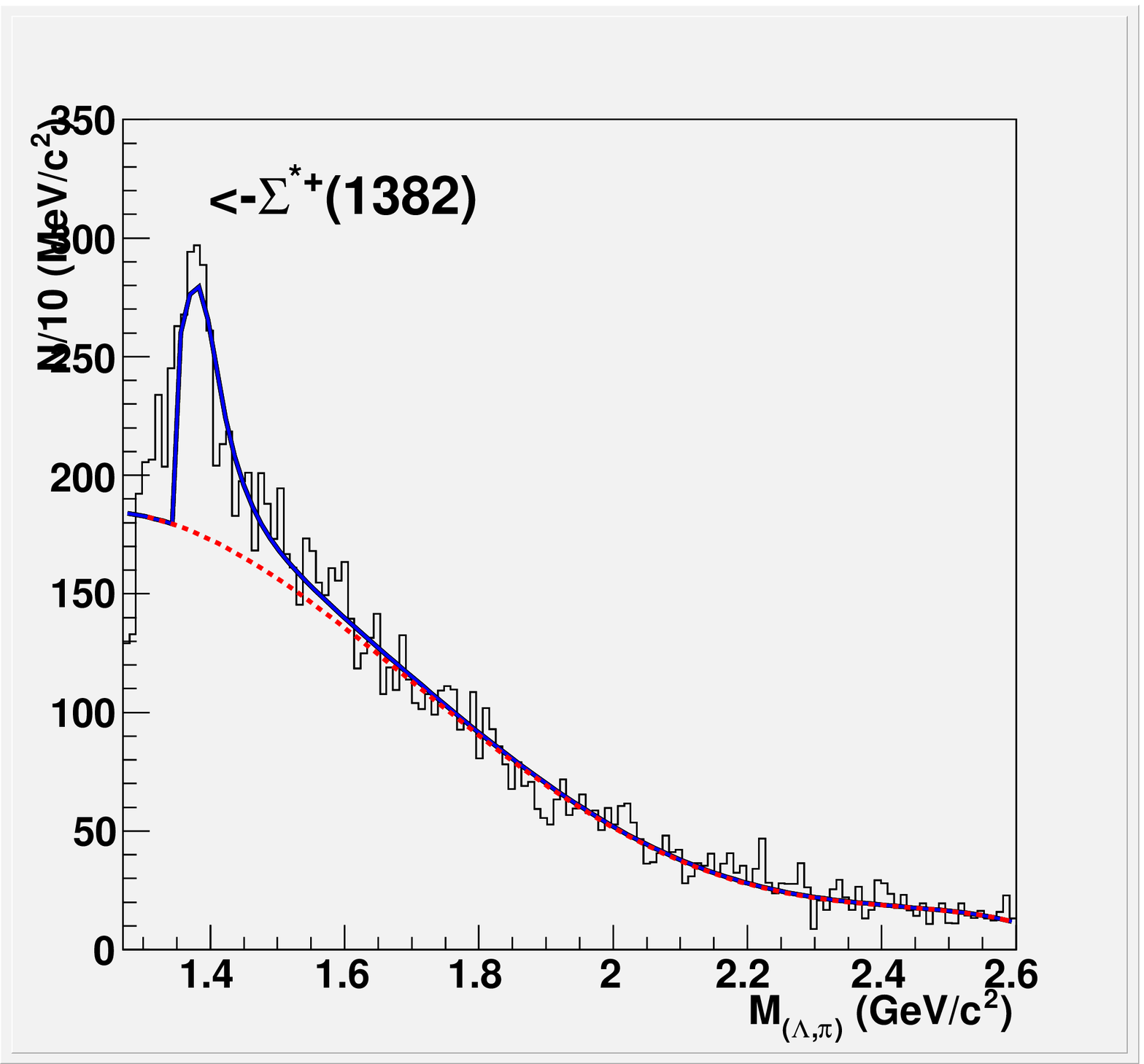}{a)}
\includegraphics[width=45mm,height=70mm]{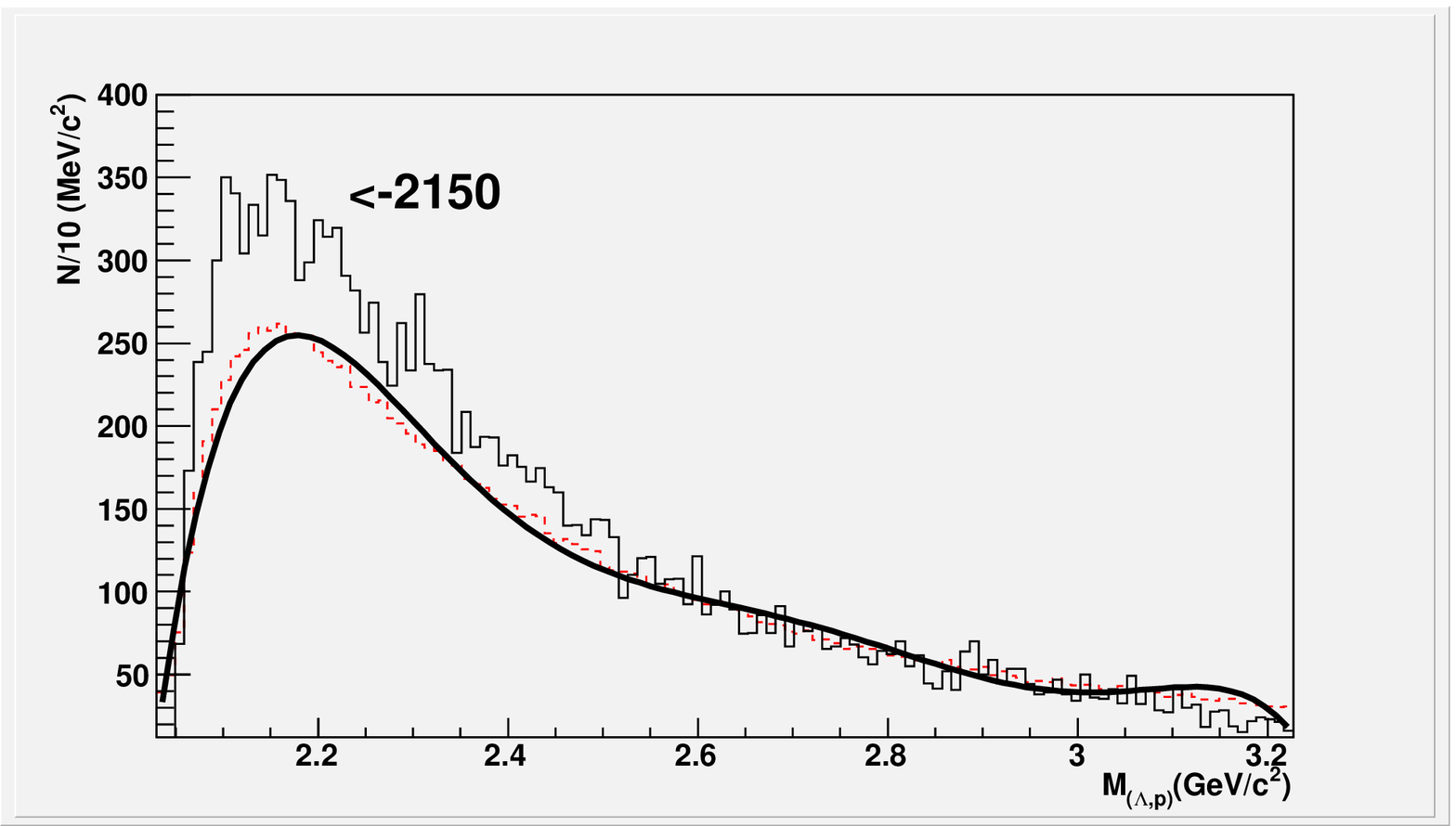}{b)}
\includegraphics[width=45mm,height=70mm]{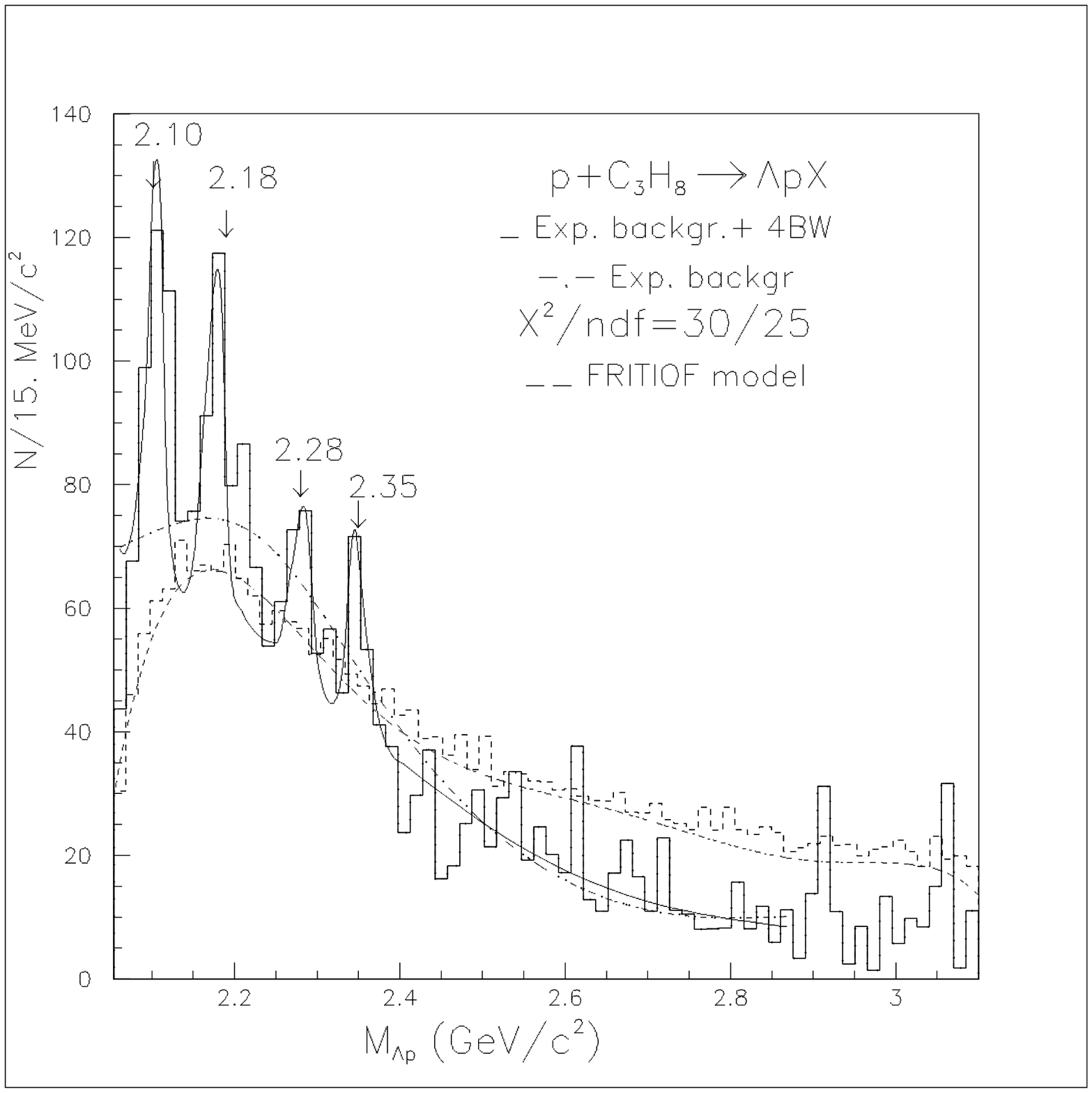}{c)}
 \caption{a)The $\Lambda \pi^+$ - spectrum;b) All $\Lambda$ p comb.c) $\Lambda p$
 spectrum for identified protons. The dashed histogram is simulated events by FRITIOF.}
\label{lpip}
\end{figure}

\begin{figure}[h]
\includegraphics[width=45mm,height=70mm]{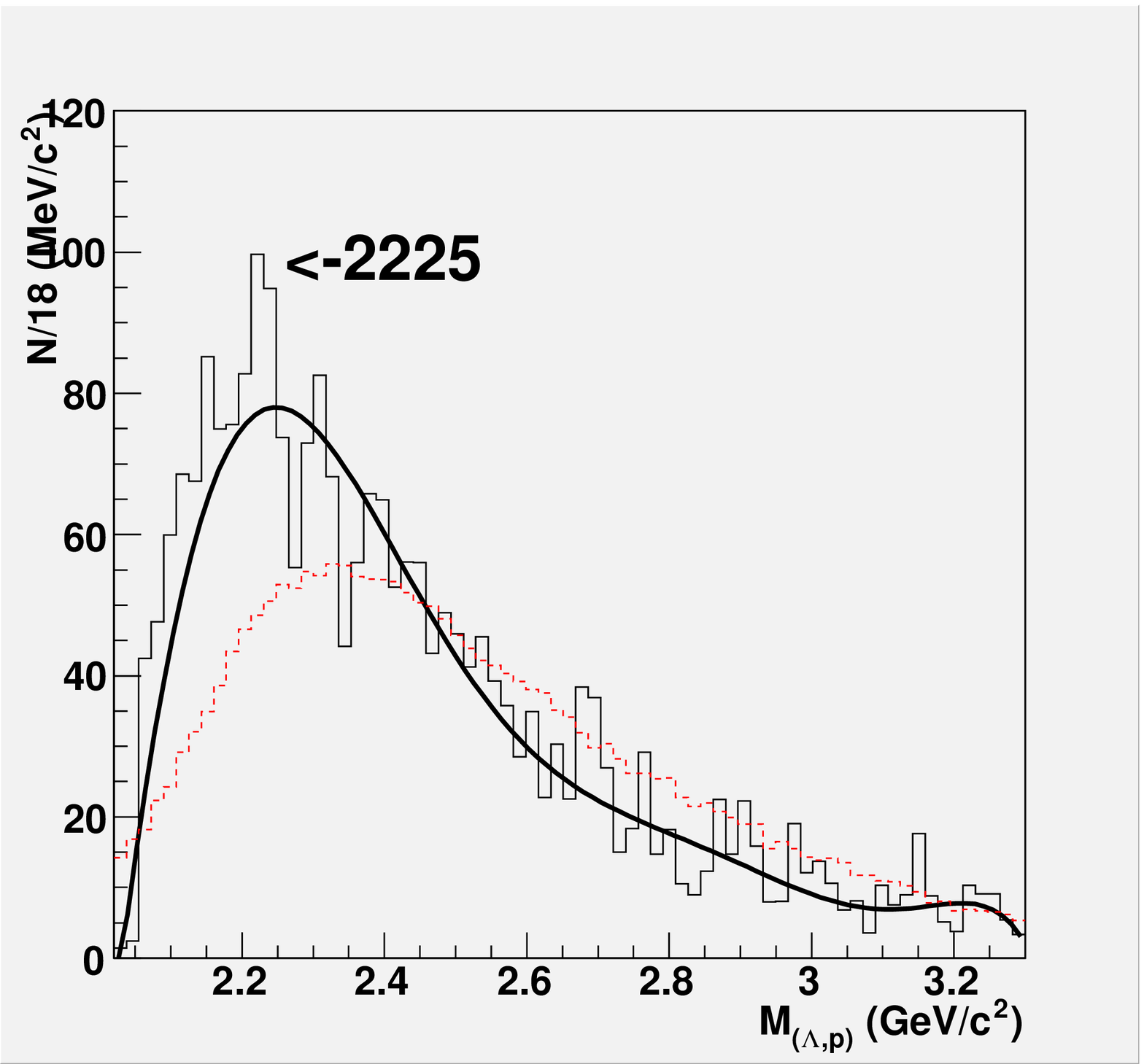}{a)}
\includegraphics[width=45mm,height=70mm]{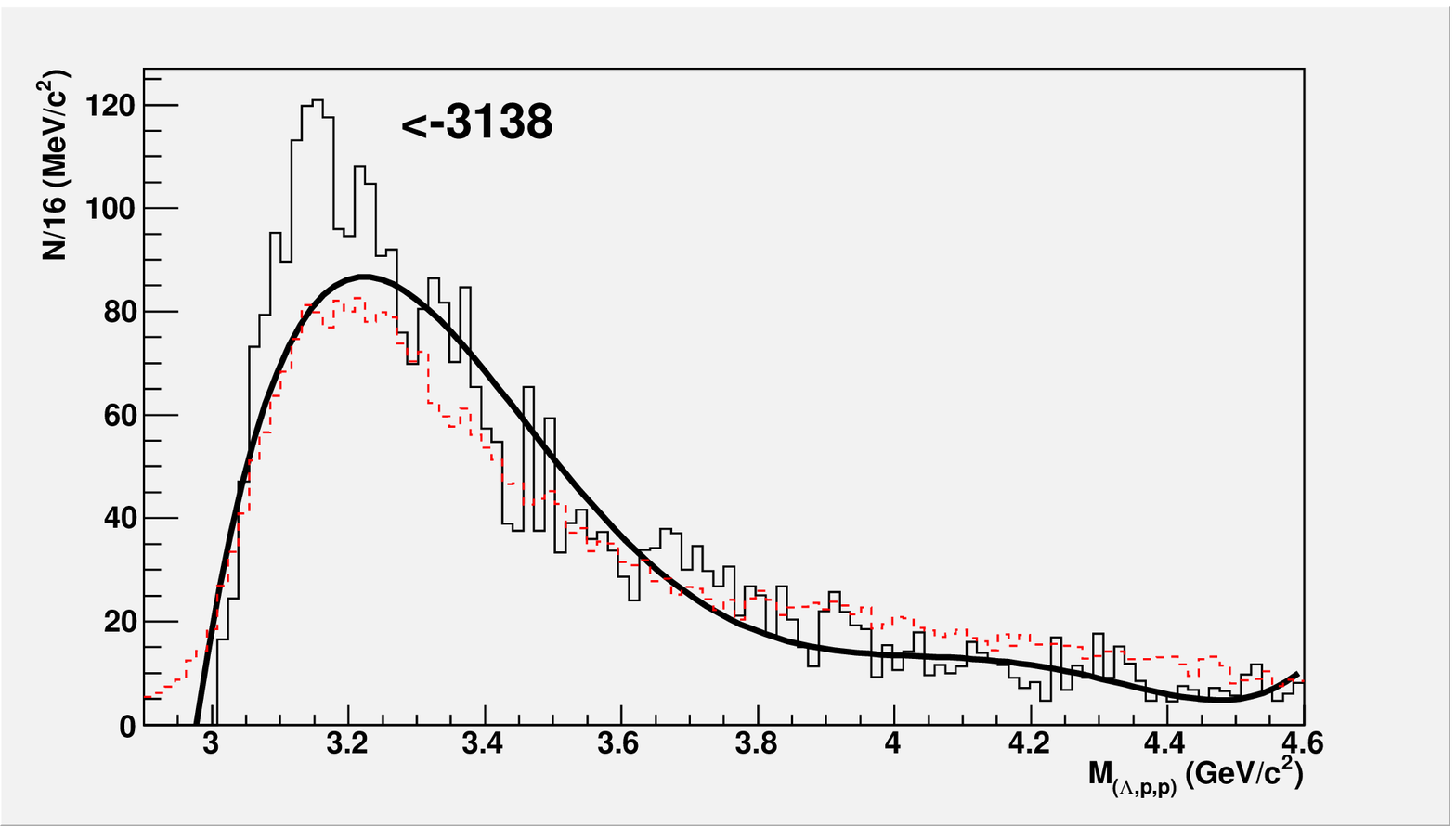}{b)}
\includegraphics[width=45mm,height=70mm]{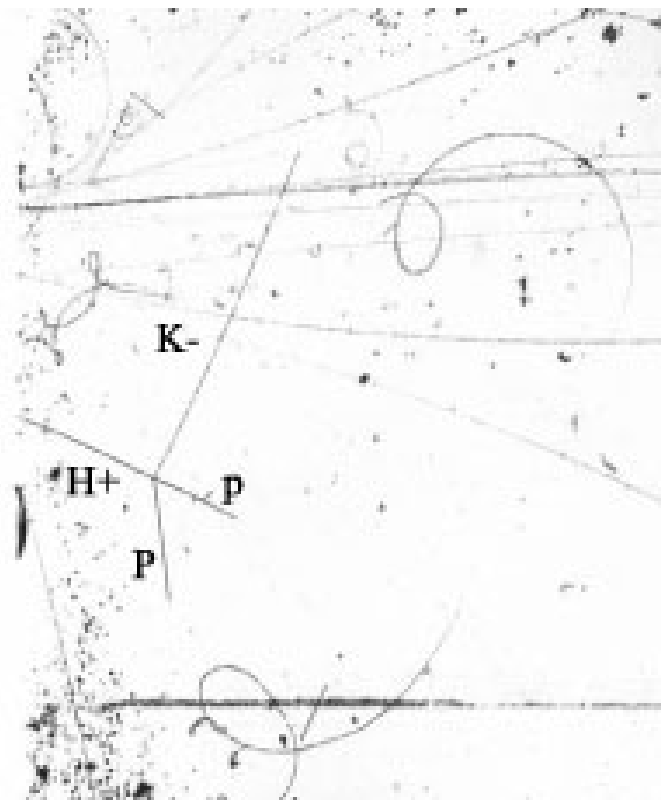}{c)}
  \caption{a)$\Lambda p$  spectrum for relativistic protons;b)
 $\Lambda p p$ spectrum for identified protons;c) The weak decay for
 $H^+ \to K^- pp$.The dashed histogram is simulated events by
 FRITIOF.
 }
 \label{lpa}
\end{figure}

 Figure  \ref{lpip}b  shows  the invariant mass of all $\Lambda
p$ combinations(13103) with bin sizes 10 MeV/$c^2
$(\cite{lp})without cuts. There are small enhancements in mass
regions of 2100, 2150, 2225 and 2353 MeV/$c^2$(Fig.\ref{lpip}b).
 Figure \ref{lpip}c  shows  the invariant mass of 2434 ($\Lambda p$)combinations
 with bin sizes 15 MeV/$c^2 $(\cite{lp})for identified protons with
 cut of momentum of 0.250$< P_p<$ 0.900 GeV/c and proton multiplicity $n_p\le$2. The values for
 the mean position  of the peak and the width obtained by using Breit Wigner fits.
  There are significant enhancements in mass regions of 2100, 2175, 2285 and 2353
MeV/$c^2$(Fig.\ref{lpip}c).Their excess above background by the
second method is 6.9, 4.9, 3.8 and 2.9 S.D., respectively. There is
also a small peak in 2225( 2.2 s.d.) MeV/$c^2$ mass region\cite{lp}.
The peak of 2180 MeV/$c^2$ was agreed with the peak from reports of
FOPI collaboration. The $\Lambda p$ effective mass distribution for
2025 combinations with relativistic protons over a momentum of P
$>$1.65 GeV/c is shown in Figure \ref{lpa}a . The solid curve is the
6-order polynomial function($\chi^2$/n.d.f=205/73). Backgrounds for
analysis of the experimental data are based on FRITIOF and the
polynomial method. There are significant enhancements in mass
regions of 2155(2.6 S.D.), 2225(4.7 S.D., with $\Gamma$=23
MeV/$c^2$), 2280(4.2 S.D.), 2363(3.6 S.D.) and 2650 MeV/c$^2$(3.7
S.D.). These peaks with relativistic protons  and with identified
protons are conformed.

The $\Lambda pp$   effective  mass distribution for 3401
combinations for identified protons with a momentum of  P $<$0.9
GeV/c is shown in Figure \ref{lpa}b. The solid curve is the  6-order
polynomial function($\chi^2$/n.d.f=245/58, Fig.\ref{lpa}b ). The
backgrounds for analysis of the experimental data are based on
FRITIOF and the polynomial method. There are significant
enhancements in mass regions of 3138(6.1 S.D.) and  3320(5.1 S.D.).
There are small enhancements in mass regions of 3087(2.2 S.D.),
3199(3.3 S.D.), 3440(3.9 S.D) and 3652MeV/$c^2$(2.6 S.D.). These
peaks in ranges of 3138 and 3199 MeV/$c^2$ were agreed with
registered peaks from reports of E471 experiment, PS,KEK.

\section{New  observation for heavy S=-2,
 $H^+\to K^-pp$ dibaryon }

Searches for stable S=-2 dibaryon states are going on\cite{H}(Table
\ref{effh}) . New candidates for S=-2 $H^+$ dibaryon shows in Fig.
\ref{lpa}c.The appearance of its first part, 15.8 cm long, with a
momentum of $p_{H^+} =1.2 \pm 0.12$GeV/c and average relative
ionization more than $I/I_0>$2 . The second part is due to two
stopped protons. The momentum of negative $K^-$ is equal to
$0.56\pm$0.03 GeV/c($I/I_0\approx$ 1.5 ). The kinematic threshold
does not permit ($\sqrt{s}$=1.96 GeV/c) imitating the reaction with
deuteron including fermi motion. The $H^+\to K^-pp$ hypothesis  fits
the event with $\chi^2$(1V-3C)=2.6, C.L.= 28\%, and $M_{H^+}$
=2482$\pm$48 MeV/$c^2$. There is also possibility for fit  by
hypothesis with decay channel $H^+\to \Sigma^+\pi^-p$ which have
much less probability than above hypothesis.

\section{Conclusion}

$\bullet{}$The experimental ratio for average multiplicities  of
$\Lambda /\pi^+ $ in the pC reaction is approximately two times
larger than this ratio from pp reactions or from simulated pC
reactions by FRITIOF model at momentum 10 GeV/c\cite{v0}. \\
$\bullet{}$The invariant mass of  $\Lambda \pi^+$  spectra has
observed significant enhancement in invariant mass range of 1382
Mev/$c^2$ ($\Sigma^{*+}$ from PDG)as a test for this method.\\
$\bullet{}$A number of important peculiarities   were observed  in
pA$\to\Lambda $p X reactions  in the effective mass spectra  for
exotic  states with decay modes:$\Lambda p$ and $\Lambda p p$.\\
$\bullet{}$A few events were registered by hypothesis of S=-2 light
$H^0$ and heavy $H^{0,+}$ dibaryons by weak decay channels  to
($\Sigma^-$,p), ($\Lambda, p, \pi^0$), ($\Sigma^+p\pi^-$), ($K^-$,p,
p).

\begin{table}
 \caption{Mass and weak decay channels for the
registrated dibaryons. } \label{effh}
%\tabcolsep7pt
%\begin{tabular}{@{}crrrr@{}}
\begin{tabular}{|c|c|c|c|c|c|c|}  \hline
Channel of decay& Mass H&C.L.&References \\
&$(MeV/c^2)$&of fit& \\
&Dibaryon&$\%$ &\\ \hline $H^0 \to \Sigma^-p$&$2172\pm15$&$
99$&Z.Phys.C  39, 151(1988).
\\   \hline $H^0 \to \Sigma^-p,\Sigma^-\to n\pi^-$&$2146\pm1$&
30& JINR RC,\\ $H^0_1 \to H^0(2146)\gamma$&$2203\pm6$& 51&N
1(69)-95-61,1995. \\   \hline $H^0 \to \Sigma^-p,\Sigma^-\to
n\pi^-$&$2218\pm12$& 69&Phys.Lett B235(1990),208. \\   \hline $H^0
\to \Sigma^-p,\Sigma^-\to n\pi^-$&$2385\pm31$&34&Phys.Lett
B316(1993),593.\\     \hline $H^+ \to p\pi^0\Lambda ^0,\Lambda^0\to
p\pi^-$&$2376\pm10$&87&Phys.Lett B316(1993),593
\\\hline $H^+ \to p\pi^0\Lambda ^0,\Lambda^0\to p\pi^-$
&$2580\pm108$&86&Nucl.Phys.75B(1999),63. \\
$H^+n \to \Sigma^ +\Lambda ^0~n, \Lambda^0\to p\pi^-$
&$2410\pm90$&6&\\ \hline $H^+ \to p\gamma\Lambda ^0,\Lambda^0\to
p\pi^-$ &$2448\pm47$&73&JINR Com.(2002)\\
$H^+ \to p\pi^0\Lambda^0,\Lambda^0\to p\pi^-$ &$2488\pm48$&72&E1-2001-265 \\ \hline
\end{tabular}

\end{table}

\end{document}